\documentclass[amssymb,aps,prl,twocolumn,epsfig,amstex]{revtex4}
\input epsf
\usepackage{graphicx} 
\usepackage{epsfig}
\usepackage{dcolumn} 
\usepackage{bm}
\usepackage{color}
\usepackage{times}
\usepackage{hyperref}
\usepackage{eurosym}
\usepackage{amsmath}
\begin{document}
\color{black}

\title{An error accounting algorithm for electron counting experiments}
\author{Michael Wulf}
\email{michael.wulf@ptb.de} \affiliation{Physikalisch Technische
Bundesanstalt, Bundesallee 100, 38116 Braunschweig, Germany}

\date{\today}

\begin{abstract}

Electron counting experiments attempt to provide a current of a
known number of electrons per unit time. We propose architectures
utilizing a few readily available electron-pumps or turnstiles with
modest error rates of 1 part per $10^4$ with common sensitive
electrometers to achieve the desirable accuracy of 1 part in $10^8$.
This is achieved not by counting all transferred electrons but by
counting only the errors of individual devices; these are less
frequent and therefore readily recognized and accounted for. Our
proposal thereby eases the route towards quantum based standards for
current and capacitance.
\end{abstract}

 \maketitle

\begin{section}{I. Introduction}
The long pursued goal of closing the quantum metrological triangle \cite{met-tri,gallop05,pekReview2012} involves the comparison of a current given by Josephson and Quantum-Hall effect, with one
based on counting electrons flowing through a constraint.
An equivalent alternative is to charge a capacitor with a counted number of electrons \cite{Keller99}.

For metrologically precise measurements of current about 1 nA is required
\cite{Keller2000}. This can be provided either by pumping at frequencies higher than those that have shown satisfactory accuracy or by
parallelization. Higher operation frequencies
increase the error-rates, while
parallelization is hindered by conflicting requirements on
the devices, including yield and accuracy. We propose here a concept
to alleviate both concerns; this is made possible by the relative simplicity of non-adiabatic single gate devices providing quantized current, which have been realized in several technologies \cite{delftQcurr,Kaestner07,Kaestner08a,pekolaCurr} and even parallelized \cite{phillip2010,parallelSINIS}

On the other hand error-rates akin to those of the shuttling experiment conducted for adiabatic pumps \cite{Keller2000,Keller99} have not been demonstrated, while direct current measurements indicate that error rates near $\Gamma=1$ ppm can be achieved \cite{Giblin2012}.

To improve upon these results, we propose a circuit that utilizes the high precision and
accuracy provided by electron pumps, but permits the correction
of the remaining errors: We connect several pumps in series and observe the charges on memory nodes between them. An example of such a circuit is shown in
fig (1).
\begin{figure}[b]\label{pic:B2A:sem}
    \centering
    \includegraphics[width=3.5in, height=1.8in]{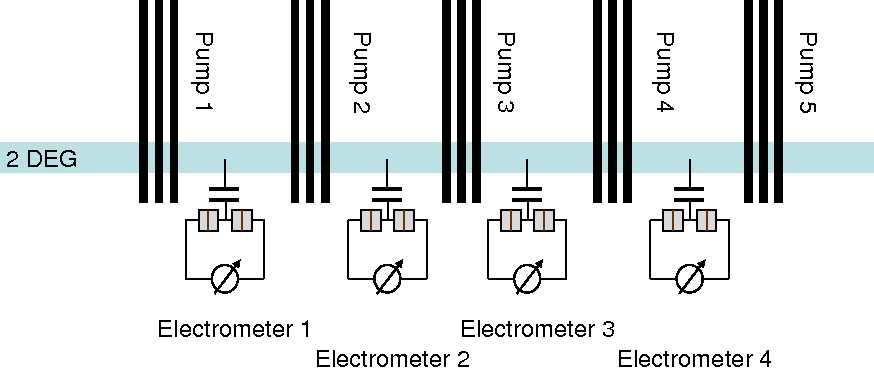}
   \caption{Schematic of a circuit for an error-correction scheme in a typical electron counting experiment composed of metallic single electron transistors and electron pumps utilizing dynamic quantum dots in a 2-dimensional electron gas.}
\end{figure}

Each of the aforementioned pumps will provide a current very close to $e\Omega$, when its gate is driven with periodicity $1/\Omega$, $-e$ being the electron charge; and will barely pass electrons between the terminals in a suitable off state.
\\
When all turnstiles move electrons with the same
rate, then only each node-charge averaged over a pump-period is constant. Whenever an error,
a surplus (or deficit) electron being shifted by pump $j$, occurs,
then the charge on node $j-1$ will drop by one electron, while the
charge on node $j$ will rise by it, labeling the
outside reservoirs as nodes $0$ and $N$. Accordingly, errors committed by different pumps result
in different charge signals on the nodes.
Such different signatures can be
recognized and attributed. If this were to be done flawlessly, it
would be possible to tell with certainty by how many electrons the
charge transported by each pump deviates from the intended
value. \\
But when the characteristic time to identify the charge state, $\tau_M$, and the time between  errors, $1/\Gamma\Omega$,
are comparable
 then, if
all observed errors are attributed to the pump (or pumps) whose
error would most likely explain the observed signature, some
multi-pump errors will mistakenly be mis-attributed:
Events, where the first two pumps of a three pump circuit both pump one electron
more than intended, cannot be distinguished from the -more likely- case of one too
few electrons pumped by the third pump. Only a four stage circuit would be able
to distinguish these two scenarios. Circuits with an odd number of pumps $N$ exhibit
$2N!/((N-1)/2)!((N+1)/2)!$ different scenarios of errors of
$(N+1)/2$ pumps that cannot be distinguished from errors of all
other pumps. If these happen within a
time-interval $\tau_M$, then they will be mis-attributed.
The relative rate of such mis-attributions is then approximately
\begin{equation}\label{eq:probabilities_simple}
\Gamma_{corrected}^{N}\simeq2\frac{N!}{\frac{N-1}{2}!\frac{N+1}{2}!}(\Gamma\Omega\tau_M)^{\frac{N+1}{2}}\frac{1}{\Omega\tau_M}.
\end{equation}
So the corrected error-rate scales as the bare
error rate, $\Gamma$, risen to the power of half the number of stages employed; this allows devices with a sufficiently low bare
error-rate to reach any desired corrected
error-rate with additional stages; this assertion is demonstrated numerically in section (III) of this paper after the required analytical framework is provided in section (II). \\

\end{section}
\begin{section}{II. Master Equation Treatment of the arbitrary error-correction circuit}

We identify the net errors the $N$ pumps have committed in time
interval $[0,t)$ with the stochastic process $\mathfrak{K}$ with
outcome space $\mathbb{Z}^N$ over time and denote the
accompanying probability distribution over $\mathbb{Z}^N$ with
$\vec{\phi}(t)$. With $\vec{i}=(i_1,i_2,...i_N)$ the entries of
vector $\vec{\phi}$ are labeled $\phi(\vec{i})(t)$ and are the joint
probabilities of $i_j$ errors having been being committed by pump
$j$ for any $j$ in the time-interval $[0,t)$. 
As the error-events we intend to observe are assumed to be rare, we
can assume $\mathfrak{K}$ to be a Markov-process. For the discrete time series
at integer multiples of a small time $\delta t$ process $\mathfrak{K}$ is described by the Pauli-Master equation
\begin{equation}\label{eq:discrete:master}
\vec{\phi}(t+\delta t)-\vec{\phi}(t)=K\vec{\phi}(t)\delta t,
\end{equation}
where $K$ is the stochastic matrix describing $\mathfrak{K}$ and
constant if $\mathfrak{K}$ is stationary.
$K$ is composed of the probabilities of an error $\vec{F}$
occurring during a time-interval $[k\delta t, (k+1)\delta t)$ given
initial state of the errors $\vec{i}$ as $P_{\vec{F}}^{\delta t}(k
\delta t,\vec{i})=\delta t K_{\vec{i}+\vec{F},\vec{i}}$. The
probability for the realization of a particular trajectory
$\mathcal{F}$, with $\vec{F}(k\delta t)$ errors committed in each
time interval $[k\delta t,(k+1)\delta t)$ is then
\begin{equation}\label{eq:piF}
\pi_\mathcal{F}=\prod_{k=0}^{T/\delta t} P_{\vec{F}(k\delta t)}^{\delta t}(k
\delta t,\sum_{k'=0}^{k-1}\vec{F}(k'\delta t))
\end{equation}
\\
In addition to the random-walk of eq. (\ref{eq:piF}) we utilize a
number $s$ of noisy measurements in the circuits of interest. For
each time the outcome of the $s$ measurements can be written as a
vector $\vec{S}$ in $\mathbb{R}^s$.
We assume that, given state $\vec{i}$ of the random-walk the
measurement outcomes follow the form
\begin{equation}\label{eq:electrometeroutputfunction}
\vec{S}_{[\vec{i}]}(k\delta t)=\vec{f}(\vec{i}(k\delta t),k\delta t)+\vec{n}(k\delta t),
\end{equation}
where $\vec{n}(t)$ is stochastic additive noise and $\vec{f}$ the
deterministic transfer-function relating observed system to measurement outcomes. We assume the noise to be independent of and uncorrelated to the random walk, to be stationary, ergodic and markov and, crucially, to be uncorrelated in time, which
is a valid approximation, when the errors are rare
compared to the noise's auto-correlation time.
For the noise
averaged over interval of length
$\delta t$ we denote the noise probability-density over $\mathbb{R}^s$
with $\vec{\nu}_{\delta t}(t,\vec{n})$.
Equivalently to the time-trajectories of the error-scenarios
$\mathcal{F}$, we define $\mathcal{S}$ to be time-trajectories of the
measurement outcomes
%
and note that the probability that the circuit commits a particular
random walk $\mathcal{F}$ and generates electrometer outputs
$\mathcal{S}$ over interval $[0,T)$ is
\begin{equation}\label{eq:PiSF}
%
\Pi_\mathcal{S}^\mathcal{F}=\pi_\mathcal{F} \prod_{k=0}^{T/\delta t}
\vec{\nu}_{\delta t}(k \delta t,\vec{S}(k \delta t)-\vec{f}(\sum_{k'=0}^k\vec{F}(k \delta t),k \delta t)),
\end{equation}
By Bayes theorem the conditional probability $\Pi_{F|S}$ of a given
random walk $\mathcal{F}$ having occurred when output $\mathcal{S}$
is exhibited is
\begin{equation}\label{eq:piSgivenF}
\Pi_{\mathcal{F}|\mathcal{S}}=\Pi_\mathcal{S}^{\mathcal{F}}/\sum_{\mathcal{F}'}\Pi_\mathcal{S}^{\mathcal{F}'},
\end{equation}
whenever the denominator is non-zero. Accordingly, given
$\mathcal{S}$ (over the entire time interval $[0,T)$) the
probability that net errors $\vec{E}$ have occurred in the
time-interval $[0,t)$ is the sum of these conditional probabilities
over all possible paths $\mathcal{F}$ satisfying said condition :
\begin{equation}\label{eq:PivecES}
\Pi(\vec{E},t|\mathcal{S})=\sum_{\mathcal{F}|\vec{E},t}\Pi_\mathcal{S}^{\mathcal{F}}/\sum_{\mathcal{F}'}\Pi_\mathcal{S}^{\mathcal{F}'}.
\end{equation}
While the probability of $\vec{E}$ errors having occurred in time
interval $[0,t)$, given measurement $\mathcal{S}$ over that time
interval and any measurement outside of it, is
\begin{equation}\label{eq:PivecES}
\Pi(\vec{E},t|\mathcal{S},t)=\frac{\sum_{\mathcal{S}'(t')=\mathcal{S}(t') \forall t'\leq
t}\sum_{\mathcal{F}'|\vec{E},t}\Pi_{\mathcal{S}'}^{\mathcal{F}'}} {\sum_{\mathcal{S}'(t')=\mathcal{S}(t') \forall
t'\leq t}\sum_\mathcal{F}\Pi_{\mathcal{S}'}^\mathcal{F}}.
\end{equation}
Where the interest is not in identifying one particular
random walk per se, but to identify the probabilities that the final
result of the underlying random walk has been $\vec{E}$ net-errors,
provided measurements $\mathcal{S}$; in particular if there is no need (or
possibility) to identify in what precise time interval an error
occurred, then indeed $\Pi(\vec{E},T|\mathcal{S},T)$ is going to be the
variable of interest. Obviously no more can be desired, once $\delta t$ is
chosen sufficiently small. \\
More convenient to compute is the probability that up to time $t$
$\vec{E}$ net errors have occurred and measurement-outcome
$\mathcal{S}$ been observed over the same interval, but irrespective
of $\mathcal{S}$ for times $t'>t$:
\begin{equation}
\psi(\vec{E},t|\mathcal{S},t)=\sum_{\vec{\mathcal{S}}'(t')=\vec{\mathcal{S}}(t') \forall t'\leq
t}\sum_{\mathcal{F}'|\vec{E},t}\Pi_{\mathcal{S}'}^{\mathcal{F}'}.
\end{equation}
When stepping forward in time by $\delta t$, the change of
$\psi(\vec{E},t|S,t)$ then includes four terms:  random walks
leaving the second sum, because of an error-event, $\vec{F}(t+\delta t)\neq0$; random walks entering the second sum for the
same reason, $\vec{F}(t+\delta t)\neq0$ and $\sum_{t'\leq
t+\delta t}\vec{F}(t)=\vec{E}$ and $\vec{S}'(t')=\vec{S}(t') \forall
t'\leq t+dt$; and paths $\mathcal{S}'$ leaving the first sum, because
$\vec{S}'(t+\delta t)\neq \vec{S}(t+\delta t)$, but satisfying
$\vec{F}(t+\delta t)=0$, to avoid double counting with respect to the
first term. Of course there are as well paths that stay within both
sums, with $\sum_{t'\leq t}\vec{F}(t)=\sum_{t'\leq
t+dt}\vec{F}(t)=\vec{E}$, and $\vec{S}'(t')=\vec{S}(t') \forall
t'\leq t+\delta t$;
while no paths can enter the first sum by its definition. \\
Accordingly
\begin{widetext}
\begin{eqnarray}\label{eq:differenceeqlong}
\psi(\vec{E},(k+1)\delta t|S,(k+1)\delta t)-\psi(\vec{E},k \delta t|S,k \delta t) =
-\sum_{\vec{F}(k \delta t)\neq\vec{0}}P_{\vec{F}}(k \delta t,\vec{E})
\psi(\vec{E},k\delta t|S,k\delta t) \nonumber\\
+\vec{\nu}(\vec{S}((k+1)\delta t)-\vec{f}(\vec{E},(k+1)\delta t))
\sum_{\vec{F}(k \delta t)\neq\vec{0}}P_{\vec{F}}(k \delta t,\vec{E}-\vec{F})
\psi(\vec{E}-\vec{F},k \delta t|S,k \delta t) \nonumber\\
-[1-\vec{\nu}(\vec{S}((k+1)\delta t)-\vec{f}(\vec{E},(k+1)\delta t))]P_{\vec{0}}\psi(\vec{E},t|S,t)\
%
%
\end{eqnarray}
\end{widetext}
where $\vec{0}$ is the null-vector. For each time $\psi$ contains
for each trajectory $\mathcal{S}$ a number of elements equal to the
number of possible net error scenarios $\vec{E}$. For a given trajectory
with $N$ pumps under consideration, at each time the set of all
$\psi(\vec{E},t,\mathcal{S},t)$ will be an element of $[0,1]^N$ 
Analogous to $\vec{\phi}$ it is convenient to summarize all
$\psi(\vec{E},t,\mathcal{S},t)$ into $\vec{\psi}(t)$. Dropping the reference
to $\mathcal{S}$. Eq. (\ref{eq:differenceeqlong}) simplifies to
\begin{equation}\label{eq:differenceeqoperator}
\vec{\psi}(t+\delta t)=(1+U\delta t)(1+K \delta t)\vec{\psi}(t) \simeq e^{U\delta t}(1+K\delta t)\vec{\psi}(t).
\end{equation}
Here $U \delta t$ includes the dependence on measurements $\vec{\nu}$ of
eq. (\ref{eq:differenceeqlong}) and is a diagonal matrix;
while $K$
does not depend on $S$ and is constant if the error-rates do not
depend on time, that is if $\mathfrak{K}$ is stationary. Remarkably the evolution of $\vec{\psi}$ is governed by a linear difference-equation akin to eq. (\ref{eq:discrete:master}). While the definitions both $P_{\vec{F}}$ and $\vec{\nu}$ depend
on the choice of $\delta t$, we assume that it can be chosen sufficiently
small, such that neither $K$ nor $U$ do.
\\
For all times $t>0$, it is obvious that
\begin{equation}
\|\vec{\psi}(t)\|_1=\sum_{\vec{E}}\psi(\vec{E},t|\mathcal{S},t)=\sum_{S'(t')=S(t')
\forall t'\leq t}\sum_F\Pi_{\mathcal{S}'}^F,
\end{equation}
so it is convenient to define that
\begin{equation}\label{eq:initialcondition}
\psi(\vec{0},0,\mathcal{S},0)=\sum_{\vec{E}}\sum_{\mathcal{S}}\psi(\vec{E},0,\mathcal{S},0)=1,
\end{equation} which implies that the set of all possible measurement
outcomes $S$ prior to the first measurement has one element of
probability $1$. With initial condition eq.
(\ref{eq:initialcondition}) and time development of eq.
(\ref{eq:differenceeqlong}) or eq. (\ref{eq:differenceeqoperator})
the final probability of $\vec{E}$ errors having occurred in
interval $[0,T)$ and trajectories $S$ having been measured can
readily be computed as $\psi(\vec{E},T,\mathcal{S},T)$; the conditional
probability of $\vec{E}$ errors having occurred in interval $[0,T)$
given trajectories $S$ having been measured is then
$\psi(\vec{E},T,\mathcal{S},T)/\sum_{\vec{E'}}\psi(\vec{E'},T,\mathcal{S},T)$, where the
denominator is just the 1-norm of vector $\psi(T)$. Eq.
(\ref{eq:differenceeqoperator}) is the master-equation governing the
net-error probabilities of the shrinking set satisfying trajectories
$S$.
\\
The expectation value and variance of the errors given trajectory
$S$ is then computed as
\begin{eqnarray}\label{eq:expectEvec}
<\vec{E}|S>=\sum_{\vec{E}}\frac{\vec{E}\psi(\vec{E},T,\mathcal{S},T)}{\sum_{\vec{E'}}\psi(\vec{E'},T,\mathcal{S},T)}
\nonumber \\
\sigma_{\vec{E}|S}=\sum_{\vec{E}}(\vec{E}-<\vec{E}|\mathcal{S}>)^2\frac{\psi(\vec{E},T,\mathcal{S},T)}{\sum_{\vec{E'}}\psi(\vec{E'},T,\mathcal{S},T)}.
\end{eqnarray}
Given measurement outcome $\mathcal{S}$ this computes that (on average)
$<\vec{E}|S>$ net-errors have occurred in time-interval $[0,T]$;
that quantity has an uncertainty of $\sigma_{\vec{E}|\mathcal{S}}$, however
the entire probability distribution $\vec{\psi}(T)$ is known. The
mean of that latter quantity weighted over all possible trajectories
will be of particular interest below,
$
<\sigma_{\vec{E}|\mathcal{S}}>_\mathcal{S}=\sum_\mathcal{S}\Pi_\mathcal{S}\sigma_{\vec{E}^\mathcal{S}},
$
the expectation-value of the uncertainty of the
error-accounting scheme. For
computer-generated $\mathcal{S}$ and $\mathcal{F}$ the average error of the
error-accounting scheme can be computed as well, as the difference,
$\delta_{\vec{E}}^\mathcal{S}$, of the computed $<\vec{E}|S>$ and ex-ante
known underlying error of the random walk. However,
$<\sigma_{\vec{E}|\mathcal{S}}>_\mathcal{S}$ is a sufficient proxy, that is accessible when only $\mathcal{S}$ is known, as in an experiment.
\\
For any random walk governed by a known stochastic matrix $K$ and
additive totally-random noise described by $U$, eq. (\ref{eq:differenceeqoperator})
is readily integrated for any one set of measurement outcomes. Expectation-value
of the occurred error and its uncertainty
according to eq. (\ref{eq:expectEvec}) follow. Average over all possible outcomes, however, can
only be approximated by summing over a large set of trajectories. \\
\end{section}

\begin{section}{III. Numerical evaluation of the one-dimensional pump chain}
The analytical treatment above is valid for an arbitrary
circuit, while for numerical studies we confine ourselves to the simplest
case, which is of particular experimental interest, the 1-dimensional chain of electron pumps.
%
%
In eq. (\ref{eq:electrometeroutputfunction}) this corresponds to $f_j(\vec{i}(t),t)=i_j-i_{j+1}$, where subindex $j$
indicates the j-th component of the vector of length $s=N-1$.
 The electrometer noise we assume to be
ergodic, stationary, gaussian, white. The random walk we assume
to be stationary - though pulsed operation of the pumps may be of
interest - and independent of the charges accumulated on the nodes.
This case is understood even using only the
arguments yielding eq. (\ref{eq:probabilities_simple}) and could
experimentally be realized by applying feedback voltages to the
nodes. Then the net-errors will diffuse freely for each
pump and will accordingly diverge with $\sqrt{T}$. As
the electrometers cannot distinguish states that
differ by the same number of errors for all pumps, we combine charge equivalent states when computing
$\vec{\psi}$. For free error diffusion we need to
compute $\vec{\psi}$ only over a few states around the most
likely error-scenario, typically all states that deviate by up to
three errors from the most likely error-scenario. Integration of eq. (\ref{eq:differenceeqlong}) is simplified as $K$ is both constant and invariant under translation by any
possible error. The latter is no longer the case when voltage dependence of the error-rates is included; the constant $K$ is
here only invariant under uniform translations of errors on all
pumps. Accordingly, $d\vec{\psi}/dt$ needs to be calculated over a
larger configuration space. This is the case of mesoscopic feedback \cite{Lukas2011,mesofeed}; however such charge dependence of the error-rates does not add conceptual complication to our treatment.
\begin{figure}
    \centering
    \includegraphics[width=3.5in, height=1.574in]{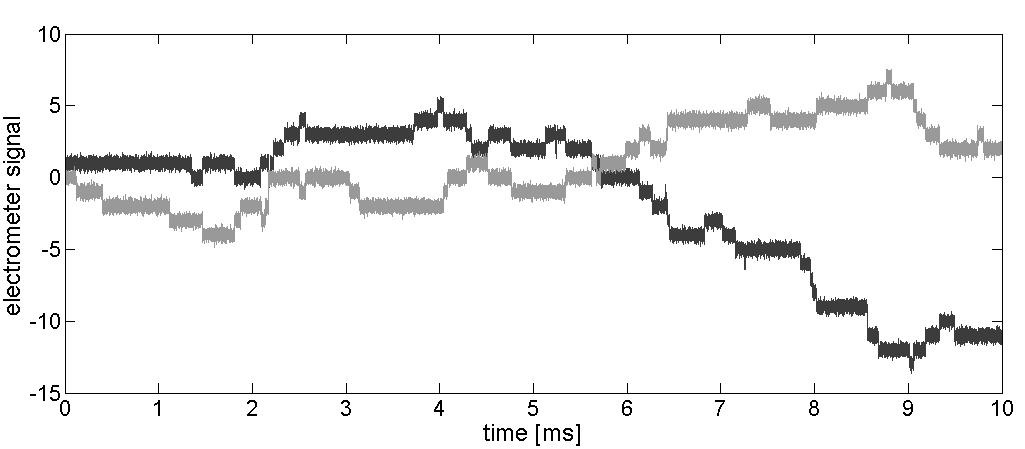}
    \caption{Signal traces of first (black) and second (gray) electrometer.
    Black and red lines indicate ideal measurement results without noise. Pump-error
    rates are $1000/$s, per pump and sign, electrometer noise $10^{-5}e/\sqrt{\textrm{Hz}}$, coupling capacitance
    ratio $C_C/C_{\textrm{node}}=0.04$.}
    \label{fig:b2a:sim:signaltraces}
\end{figure}
Figure \ref{fig:b2a:sim:signaltraces} shows a typical set of
signal traces that can be expected from the combination of the
diffusion dynamics and electrometer noise realistic for the RF-SET
\cite{RF-SET,Bylander04,ref:RFset4K} and similar devices \cite{Zorin2000,ref:AZ:LSET,ref:LSET}. These traces are used to numerically integrate eq.
(\ref{eq:differenceeqoperator}) yielding a perceived
diffusion.
The result of this perceived diffusion is subtracted from the underlying diffusion.
This difference is the error that our algorithm mis-attributes. Nonetheless, the rate
of such errors after correction is much smaller than the rate of errors of each single device.
\begin{figure}
    \centering
   \includegraphics[width=3.5in, height=1.574in]{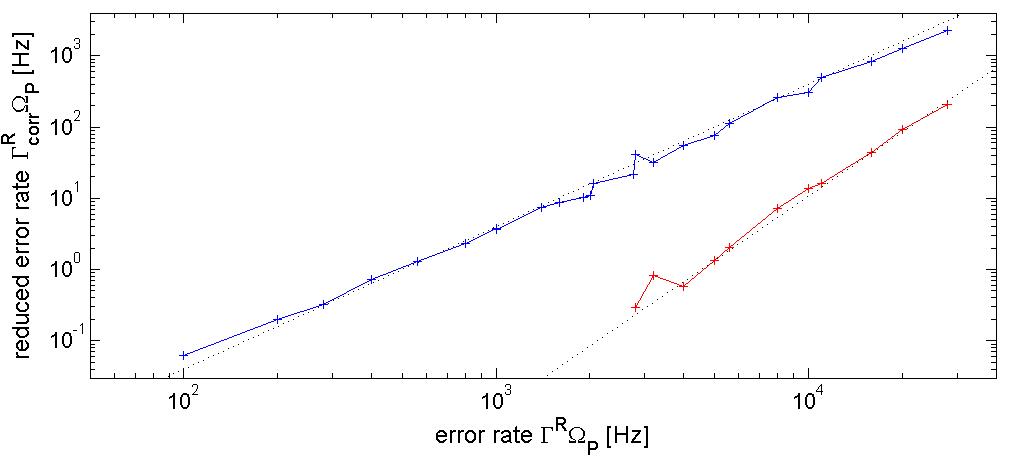}
    \caption{Corrected error rates with respect to time for three (upper curve) and five (lower curve) pump
    circuits as a function of the uncorrected error rates. Dashed
    lines are a fit to eq. (\ref{eq:probabilities_simple}). Other
    parameters as in fig. \ref{fig:b2a:sim:signaltraces}.}
    \label{fig:b2a:sim:ErrorRates}
\end{figure}
Numerical results for rates of these reduced errors are shown for various parameters in fig.
\ref{fig:b2a:sim:ErrorRates} in excellent agreement with eq.
(\ref{eq:probabilities_simple}). Even for a rate of $200$ errors per
second a $1000$-fold improvement of the error rate is achieved with
only three pumps and two intermediate electrometers, so that even
devices exhibiting a relative error-rate of $10$ ppm can reach the
metrologically desirable error threshold of $10$ ppb, due to the
proposed architecture. 
 The five pump version achieves a 10000-fold improvement
even at a rate of $3000$ errors per second. Pumps with intrinsic error-rates of $100$ ppm at pump rates of 30 MHz could thus be corrected to the required $10$ ppb accuracy. If devices with $10$ ppm were to be used, a 1000-fold improvement were to suffice, so that more than $8000$ errors per second could be accounted for. This would correspond to a pump rate of $800$ MHz, so that both the requirements of current magnitude and accuracy are met. Recall that the lowest reported error-rate for single parameter pumps is about $1$ ppm \cite{Giblin2012}.
%
%
\end{section}
\begin{section}{Conclusion}
In summary, we have proposed a family of architectures utilizing
either conventional electron pumps of modest quality or one of the
single-gate devices that are capable of producing quantized current
steps with the now common RF-SET or a similar device. The algorithm
constructed in this work determines the charge transported by such
structure to a much greater precision than would be feasible by any
single of the devices used. The relaxation on the requirements for
the quality of the individual devices used, simplifies
experimentation, even at higher operation frequency, and increases
sample yield. It thereby opens the prospects towards parallelization
needed to increase the provided current. We note that the proposed
architecture allows to monitor the error-rates of individual pumps
on the level of single electrons, similar to those stated for adiabatic
electron-pumps in their shuttle mode \cite{Keller99}. This has not been possible for
the single gate devices \cite{delftQcurr,Kaestner07,pekolaCurr} in
currently employed circuits. Accordingly the architecture proposed
here allows further research on these devices and thereby a
determination regarding their suitability for metrological
applications, while easing the required thresholds. During the preparation of this manuscript we participated in work demonstrating the integrated operation of the components required for the experimental realization of this concept \cite{ref:wulf_cpem2012}.
\end{section}

This work is supported in part by the European Union under the Joint
Research Projects REUNIAM. I gratefully acknowledges fruitful
discussions with L. Fricke, B. Kaestner, F. Hohls, R. Dolata and H-W. Schumacher
during the course of this work.

\end{document}